\documentclass[12pt]{article}
\begin{document}
\baselineskip=18 pt

\begin{center}
{\large KdV solitons in Einstein's vacuum field equations}
\end{center}

\vspace{.5cm}

\begin{center}
{\bf Debojit Sarma}\footnote{debojitsarma@yahoo.com} and 
{\bf Mahadev Patgiri}\footnote{mahadev@scientist.com} \\
{\it Department of Physics, Cotton College,}\\
{\it Guwahati-781001, India}
\end{center}

\vspace{.5cm}

\begin{abstract}
We present a metric for which Einstein's field equations in vacuum generate the 
Kortweg-de Vries (KdV) equation and thus its $N$-soliton solutions solve the vacuum equations. 
The metric of the one soliton solution has been investigated and is a non-singular, Lorentzian 
metric of type N in the Petrov classification.
\end{abstract}



\vspace{.5cm}

\section{Introduction}

Einstein's field equations are inherently nonlinear and hence exact solutions for them are not easy to obtain. 
Because of their nonlinearity, they are expected to have soliton solutions, at least for some spacetimes. 
These are pulse-like travelling wave solutions having fascinating properties, one being that they propagate 
without noticeable dispersion and interact without change of form although they do not obey linear 
superposition. Einstein's equations admit wave solutions which is 
evident on linearization of these equations for some metrics. The problem of obtaining such solutions
for the field equations without linearization is however a more difficult one, one reason being that 
no general solution of these equations is known. The techniques of extracting soliton solutions of 
nonlinear evolution equations have been applied to the task of obtaining exact solutions of Einstein's equations.
Belinski and Zakharov \cite{Bel1,Bel2,Bel3} (and \cite{Bel} for a comprehensive overview) modified 
the inverse scattering transform (IST) for application to Einstein's equation in vacuum, and since then there 
has been a consistent search for soliton solutions of the vacuum equations based on their technique. Different 
metrics are taken as the seed metric, from which soliton solutions are generated by the IST. In this manner, 
a number of solutions of the Einstein equations have been found.

In this paper, we approach the problem from a different direction. We attempt to construct a metric for which 
the vacuum Einstein's equations produce a well-known, nonlinear, evolution equation, namely the Kortweg-de Vries 
(KdV) equation \cite{KdV}, which is known to be integrable, so that the existence of $N$-soliton solutions and integrability 
of the system is assured. Substitution of any of the soliton solutions in the metric will result in a spacetime 
solving the vacuum Einstein's equations. We specifically check the metric with one soliton solution and study 
its properties, particularly, the curvature and classification scheme. The metric is Lorentzian and using the 
Peres algorithm \cite{Peres}, we find that the metric obtained here belongs to the type N in the Petrov classification 
scheme and is therefore a metric with radiation. Although the metric apparently has a singularity at $t=0$,
we show that this is not a physical singularity and appears only due to the choice of the coordinate system.

The plan of this paper is as follows: in section two, we present the metric that generates the KdV equation. In 
section three, the metric with one soliton solution is discussed in detail. The fourth section is the concluding one.

\section{Vacuum spacetime and the KdV equation}
      
We introduce the line element
\begin{equation}
ds^2=-\left[af^{2}(x,t)-2f_{xx}(x,t)\right]dt^2+2\left (\frac{3}{2}t\right )^{\frac{4}{3}}dx^2
-2f(x,t)dtdx+dy^2+2\left (\frac{3}{2}t\right )^{\frac{2}{3}}dxdy+dtdz
\label{1}
\end{equation}
where $a$ is a constant and the subscripts denote partial derivatives.
The only nonvanishing element of the Ricci tensor for the metric (\ref{1}) 
is $R_{tt}$ and the corresponding vacuum equation for it, $R_{tt}=0$ yields 
the following nonlinear equation
\begin{equation}
f_{xt}(x,t)-af(x,t)f_{xx}(x,t)-af_{x}^{2}(x,t)
+f_{xxxx}(x,t)=0.
\label{2}
\end{equation} 
Setting $a=6$ and performing one integration over $x$ leads to one of the best-known
integrable nonlinear partial differential equation in the literature, namely,
the KdV equation in its standard form (the integration constant vanishes for soliton 
solutions \cite{New}) 
\begin{equation}
f_{t}(x,t)-6f(x,t)f_{x}(x,t)+f_{xxx}(x,t)=0. 
\label{3}
\end{equation}

It is well known that the KdV equation has all the beautiful properties that 
characterize an integrable, 
nonlinear system including an infinite number of conserved quantities in involution 
(which defines Liouville integrability), $N$-soliton solutions, a bi-hamiltonian 
structure among other properties (for a review of soliton equations, specially the KdV
equation, and their 
properties see \cite{New,Ablow,Draz,Laksh}). Apart from the source of much new 
mathematics, this equation appears in a wide variety of physical problems.
In integrable systems, like the KdV, dispersion is compensated by nonlinearity 
giving rise to soliton solutions of all orders called $N$ soliton solutions. 

The one soliton solution (1SS) of the KdV equation is
\begin{equation}
f(x,t)=-\frac{k^2}{2} \mbox{sech}^2\frac{\eta}{2}
\label{4}
\end{equation}
where $\eta=kx-k^3t$, $k$ being the wave number. Equation (\ref{4}) represents a inverted pulse
of $sech^2$ profile travelling in the positive $x$ direction. 
The KdV equation being nonlinear, linear combinations of the one soliton solution 
do not provide new solutions. Instead, two soliton solutions (2SS) may be constructed, 
which represent a nonlinear interaction of two solitons. For the KdV, the  2SS is
\begin{equation}
f(x,t)=\frac{1}{2}\left (k_{1}^{2}-k_{2}^{2}\right )\left[\frac{k_{2}^{2}
\mbox{cosech}^2(\eta_{2} /2)
+k_{1}^{2}\mbox{sech}^2(\eta_{1} /2)}{(k_{2}\mbox{coth}(\eta_{2} /2)-k_{1}
\mbox{tanh}(\eta_{1} /2))^2}\right ]
\label{5}
\end{equation}
where $\eta_1=k_{1}x-k_{1}^{3}t$ and $\eta_2=k_{2}x-k_{2}^{3}t$.
That (\ref{4}) and (\ref{5}) satisfy the KdV equation (\ref{3}) can be verified
by direct substitution. 

Existence of soliton solutions alone does not imply integrability, 
although it is relevant to mention that existence of at least three soliton solutions 
is a necessary condition for it. It is well established that the KdV equation passes 
all tests of integrability.

\section{Properties of the spacetime with one soliton solution}

If we substitute $f(x,t)$ from (\ref{4}) into the metric (\ref{1}), we obtain the following
form for it.
\begin{equation}
ds^2=-k^4 \mbox{sech}^2\frac{\eta}{2}dt^2+2\left (\frac{3}{2}t\right )^{\frac{4}{3}}dx^2 
+k^2 \mbox{sech}^2\frac{\eta}{2}dtdx+dy^2+2\left (\frac{3}{2}t\right )^{\frac{2}{3}}dxdy+dtdz
\label{6}
\end{equation}
which satisfies the vacuum Einstein equations. Substituting the two soliton solution (\ref{5}) in 
(\ref{1}) would again lead to a metric satisfying the vacuum equations. In this manner, a 
hierarchy of solutions of the vacuum equations can be constructed.

The metric (\ref{6}) is Lorentzian with determinant 
$detg=-\frac{1}{4}\left(\frac{3}{2}t\right )^{\frac{4}{3}}$.
The nonzero components of the Riemann-Christoffel curvature tensor $R^{\mu}_{\nu\rho\sigma}$
 corresponding to the metric (\ref{6}) are
\begin{eqnarray}
R^1_{441}&=&-\frac{2}{9t^2} \\
R^1_{442}&=&\frac{(2/3)^{2/3}}{3t^{8/3}} \\
R^2_{441}&=&-\frac{(2/3)^{1/3}}{3t^{4/3}} \\
R^2_{442}&=&\frac{2}{9t^2} \\
R^3_{141}&=&\frac{2^{2/3}3^{1/3}}{t^{2/3}} \\
R^3_{142}&=&\frac{4(2/3)^{1/3}}{3t^{4/3}} \\
R^3_{241}&=&\frac{4(2/3)^{1/3}}{3t^{4/3}} \\
R^3_{242}&=&\frac{2}{9t^2} \\
R^3_{441}&=&\frac{2 k^2 \mbox{sech}^2(\eta/2)}{9 t^2} \\
R^3_{442}&=&\frac{(\frac{2}{3})^{2/3} k^2 \mbox{sech}^2(\eta/2)}{3 t^{8/3}}
\end{eqnarray}
It appears from the curvature components above, as well as the determinant of $g_{\mu\nu}$, 
that the metric is singular at $t=0$. However, this is merely a coordinate singularity. 
To show this, we first note that the curvature scalars $R^{\mu\nu\rho\sigma}R_{\mu\nu\rho\sigma}$ and 
$R_{\mu\nu\rho\sigma}R^{\rho\sigma\lambda\tau}R_{\lambda\tau}^{\;\;\;\;\;\mu\nu}$ do not show any singularity 
being constants equal to zero. Moreover, a coordinate transformation
\begin{equation}
t=\frac{2}{3}e^{-3\tau}
\label{6a}
\end{equation}
allows us to write (\ref{6}) as
\begin{equation}
ds^2=-4k^4e^{-6\tau} \mbox{sech}^2\frac{\xi}{2}d{\tau}^2+2e^{-4\tau}dx^2 
-2k^2e^{-3\tau} \mbox{sech}^2\frac{\xi}{2}d\tau dx+dy^2+2e^{-2\tau}dxdy+2e^{-3\tau}d\tau dz
\label{6b}
\end{equation}
where $\xi=kx-\frac{2}{3}k^3e^{-3\tau}$. The metric, in the form given by (\ref{6b}) remains Lorentzian with 
its determinant being $-e^{-10\tau}$.

For the spacetime (\ref{6b}), the components of the curvature tensor are
\begin{eqnarray}
R^1_{441}&=&-2 \\
R^1_{442}&=&-3e^{2\tau} \\
R^2_{441}&=&-2e^{-2\tau} \\
R^2_{442}&=&2 \\
R^3_{141}&=&-6e^{-\tau} \\
R^3_{142}&=&-4e^{\tau} \\
R^3_{241}&=&-4e^{\tau} \\
R^3_{242}&=&-e^{3\tau} \\
R^3_{441}&=&2 k^2 \mbox{sech}^2\frac{\xi}{2} \\
R^3_{442}&=&3e^{2\tau} k^2 \mbox{sech}^2\frac{\xi}{2}
\end{eqnarray}
The curvature components above are clearly non-singular. In addition, the curvature scalars 
$R^{\mu\nu\rho\sigma}R_{\mu\nu\rho\sigma}$ and $R_{\mu\nu\rho\sigma}R^{\rho\sigma\lambda\tau}R_{\lambda\tau}^{\;\;\;\;\;\mu\nu}$ 
are nonsingular.

Consider the one soliton  metric in form given by (\ref{6}). Under the asymptotic condition $\eta\rightarrow\infty$,
$\mbox{sech}^2\frac{\eta}{2}\rightarrow 0$, and we also find that the value of the determinant of the metric tensor does not 
change and only the curvature components $R^3_{441}$ and $R^3_{242}$ vanish, other components remaining unchanged. Hence,
the metric remains non-singular under the above-mentioned asymptotic condition. 

For vacuum solutions, the Weyl conformal tensor $C_{\mu\nu\rho\sigma}$ and the curvature tensor 
$R_{\mu\nu\rho\sigma}$ become identical. We apply the Peres algorithm to
determine the Petrov classification of the conformal tensor. In the present case, 
\begin{equation}
C_{\mu\nu\rho\sigma}\neq 0
\label{7}
\end{equation}
 however
\begin{equation}
C_2=\star C_2=0\;\;\;\;\; \mbox{and}\;\;\;\;\;\; C_3=\star C_3=0
\label{8}
\end{equation}
where  
\begin{eqnarray}
C_2&=&C^{\mu\nu\rho\sigma}C_{\mu\nu\rho\sigma} \\ 
\star C_2&=&\star C_{\mu\nu\rho\sigma}C^{\mu\nu\rho\sigma} \\
C_3&=&C_{\mu\nu\lambda\xi}C^{\lambda\xi\rho\sigma}C^{\mu\nu}_{\;\;\;\;\;\rho\sigma} \\
\star C_3&=&\star C_{\mu\nu\lambda\xi}C^{\lambda\xi\rho\sigma}C^{\mu\nu}_{\;\;\;\;\;\rho\sigma}
\end{eqnarray}
where
\begin{equation}
\star C_{\mu\nu\rho\sigma}=\frac{1}{2}\eta_{\mu\nu\lambda\xi}C^{\lambda\xi}_{\;\;\;\;\;\rho\sigma}
\end{equation}
$\eta_{\mu\nu\lambda\xi}$ being the fully anti-symmetric tensor in four dimensions. The
Peres algorithm states that equations (\ref{7}) and (\ref{8}) imply that the spacetime is 
type N or III. Additionally, in our case, the following contraction of the Weyl tensor
$C_{\mu\nu\lambda\xi}C^{\lambda\xi}_{\;\;\;\;\;\rho\sigma}=0$ which means that the metric
is of type N.
The geodesic equations for a material particle in the metric (\ref{6}) are found to be
\begin{eqnarray}
\frac{du^0}{d\tau}&=&0 \\
\frac{du^1}{d\tau}&=&-2\left [\frac{u^1u^0}{t}+\frac{\left (\frac{2}{3}\right )^{2/3}u^2u^0}{3t^{5/3}}\right ]\\
\frac{du^2}{d\tau}&=&\frac{2\left (2^{1/3}(3t)^{2/3}u^1+u^2\right )u^0}{3t} \\
\frac{du^3}{d\tau}&=&\frac{2}{9t^{5/3}}\left [k^2\mbox{sech}^2(\eta/2)\left (9t^{2/3}u^1-9t^{5/3}
k^3u^1\mbox{tanh}(\eta/2)+2^{2/3}3^{1/3}u^2\right )u^0 \right. \nonumber \\
&&+3^{1/3}(2t)^{4/3}u^1\left (3\times 2^{1/3}t^{2/3}u^1+3^{1/3}u^2\right ) \nonumber \\
&&\left.
+36t^{5/3}k^3\mbox{cosech}^3(\eta)\mbox{sinh}^4(\eta/2)\left ((u^1)^2+k^4(u^0)^2\right )\right]
\end{eqnarray}
The geodesic equations above are coupled differential equations which cannot be easily solved by
inspection. However, the symmetries of the spacetime may help in obtaining their solutions and will
be investigated in the next communication.

\section{Conclusion}

It is clear, therefore, that we have a metric for which the vacuum Einstein's 
equations reduce to the KdV equation which is known to be completely integrable. Thus the
solutions of the Einstein equations, for this metric, are the soliton solutions of the KdV 
equations. From this metric, we get the one soliton metric which is non-singular and of Petrov
type N. In this context, we note that many of the soliton solutions found by the IST do not 
have properties that characterize solitons such as permanence of form and amplitude \cite{Bel}. 
For soliton solutions generated by the metric studied in this paper, however, we expect these 
properties to be present since they are the KdV solitons. In this paper, classification and curvature
of the metric (\ref{1}) have been studied; its isometries and other symmetry properties 
will be presented in a future work.

\end{document}